\documentclass[aps,prd,notitlepage,showpacs,nofootinbib,superscriptaddress,10pt]{revtex4-1}
\usepackage{graphicx}

\usepackage{amsmath}
\usepackage{subfigure}
\usepackage{amssymb}
\usepackage{float}
\usepackage{comment}
\usepackage{slashed}
\usepackage[normalem]{ulem}

\usepackage{booktabs}
\usepackage{mathtools,braket,blkarray}

\usepackage{amsmath}
\usepackage{amssymb}

\usepackage[usenames,dvipsnames]{color}
\usepackage[colorinlistoftodos]{todonotes}
\usepackage[colorlinks=true,allcolors=darkred, pdfborder={0 0 0}]{hyperref}

\usepackage[normalem]{ulem}

\usepackage{multirow}
\definecolor{darkred}{rgb}{0.6,0,0}
\usepackage[colorinlistoftodos]{todonotes}

\definecolor{linkcolor}{rgb}{0,0,0.5}
 
\newcommand {\ignore}[1]{}

\def \znbb {$\rm 0\nu\beta\beta$}

\def\gsim{\raise0.3ex\hbox{$\;>$\kern-0.75em\raise-1.1ex\hbox{$\sim\;$}}}
\def\lsim{\raise0.3ex\hbox{$\;<$\kern-0.75em\raise-1.1ex\hbox{$\sim\;$}}}

\usepackage{soul}

\definecolor{mightnightblue}{RGB}{25,25,112}

\definecolor{brown}{rgb}{0.59, 0.29, 0.0}

\def\21{$\mathrm{SU(2)_L \otimes U(1)_Y}$}

\def\mt{$\mu$-$\tau$ reflection symmetry}

\newcommand{\AddrUNAM}{ {\it Instituto de F\'{\i}sica, Universidad Nacional Aut\'onoma de M\'exico, A.P. 20-364, Ciudad de M\'exico 01000, M\'exico.}}

\begin{document}

\bibliographystyle{unsrt}

\title{\boldmath RGE-induced $\mu$-$\tau$ symmetry breaking: an analysis of the latest T2K results }

\author{Guo-yuan Huang}\email{huanggy@ihep.ac.cn}
\affiliation{Institute of High Energy Physics, Chinese Academy of Sciences, Beijing 100049, China}
\affiliation{School of Physical Sciences, University of Chinese Academy of Sciences, Beijing 100049, China }
\author{Newton Nath}\email{newton@fisica.unam.mx}
\affiliation{\AddrUNAM}



\begin{abstract}
\vspace{0.5cm}
{\noindent
The T2K collaboration has recently reported their results, which gives the best-fit values of the atmospheric mixing angle $\sin^2 \theta_{23} = 0.53$ and the Dirac CP-violating phase $\delta = -1.89$ for normal neutrino mass ordering.  We give a possible theoretical origin of such values based on the $\mu$-$\tau$ reflection symmetry. 
It has been found that the breaking of such symmetry using  one-loop renormalization-group equations (RGEs) in the framework of minimal supersymmetric standard model can fit well with the latest T2K results and the recent global fits. 
To make a quantitative analysis, we have included  for the first time the complete dataset of oscillation, beta decay, neutrinoless double-beta decay and cosmological observations in comparing the theory to experiments.
We also further examine the importance of such breaking patterns for neutrinoless double-beta decay experiments. 
}  
\end{abstract}



\maketitle
\noindent

\section{Introduction}

The discovery of neutrino oscillations~\cite{Tanabashi:2018oca} has been confirmed by phenomenal neutrino data from solar, atmospheric, reactor, and accelerator neutrino experiments~\cite{deSalas:2017kay,Capozzi:2020qhw,Esteban:2018azc}.
The standard three-flavor neutrino oscillations are described by six parameters, namely three mixing angles $(\theta_{12}, \theta_{13}, \theta_{23})$, two mass squared differences $(\Delta m^2_{21}, \Delta m^2_{31})$ and the Dirac CP-violating phase $\delta$.
At present, the unknowns involving neutrino oscillations are the following: (i) the sign of $\Delta m^2_{31}$ ($\Delta m^2_{31} >0$ for normal neutrino mass ordering, while $\Delta m^2_{31}<0$ for the inverted one), (ii) the octant of $\theta_{23}$ ($\theta_{23} > 45^\circ$ is known as the higher octant, while $\theta_{23} < 45^\circ$ as the lower octant), and (iii) the precise value of $\delta$.
The maximal atmospheric mixing, i.e. $\theta^{}_{23} = \pi/4$, and possible maximal CP violation, i.e. $\delta = 3\pi/2$, were mildly indicated by the oscillation data, which opens up an appealing theoretical and phenomenological possibility.
The precision measurements of these unknowns are the primary goals of the upcoming neutrino  oscillations experiments. 
Moreover, the nature of neutrinos, whether they are Majorana or Dirac fermions, is not yet answered. 
If neutrinos turn out to be Majorana particles, there exist two additional Majorana CP-violating phases $\rho$ and $\sigma$ in the mixing matrix, which might be measurable via the neutrinoless double-beta decay ($0\nu\beta\beta$) experiments~\cite{Xing:2015zha,Xing:2016ymd,Ge:2016tfx,Penedo:2018kpc,Cao:2019hli,Ge:2019ldu}.

Recently, the T2K collaboration~\cite{Abe:2019vii} has published their latest measurements on $\theta^{}_{23}$ and $\delta$. Their best-fit values along with the $1\sigma$ errors can be read as
$\sin^2 \theta_{23} = 0.53^{+0.03}_{-0.04},~\delta = -1.89^{+0.70}_{-0.58}  $
for normal ordering
\footnote{It should be noted that T2K alone already shows a preference for
normal ordering with a posterior probability of 89\%. Meanwhile the global analysis of neutrino oscillation data  prefers normal ordering with more than $3\sigma$ confidence level (C.L.)~\cite{deSalas:2017kay,Capozzi:2020qhw,Esteban:2018azc}. Hence, we will focus on  normal ordering throughout this work. }.
For the first time, the experiment has ruled out $\delta$ that causes a large increase in the antineutrino oscillation probability ($\sin\delta \gtrsim 0$) at a $3\sigma$ C.L.
Moreover,
the results of latest T2K measurement indicate that $\theta^{}_{23}$ is located at the upper octant and $\delta$ is at the third quadrant, mildly deviating from $(\theta^{}_{23},\delta) = (\pi/4,3\pi/2)$ by nearly $1\sigma$ significance.
For comparison,
the results from the global analysis groups are summarized in Table~\ref{tab:Global-fit}. Note that the recent T2K data have been included in the latest global analysis of Ref.~\cite{Capozzi:2020qhw}.
%
%

Motivated by the latest T2K measurements, in this note, we plan to give a theoretical explanation  for their latest results, with a detailed quantitative calculations.
Among a large variety of theoretical models to  explain the observed leptonic mixing patterns
\footnote{In view of the so-called quark-lepton complementarity relation~\cite{Xing:2005ur}, it is interesting to notice a speculative relation $\delta - \delta^{}_{\rm CKM} = 180^{\circ}$, with $\delta^{}_{\rm CKM} \simeq 71^{\circ}$ being the CP-violating phase of the quark mixing matrix and $\delta \simeq 252^{\circ}$ takes the best-fit value of T2K. This may opens another route to explaining that $\delta$ tends to deviate from the case of maximal CP violation: the amount of CP violation is the same for quark and lepton sectors while the CP is not maximally violated in the quark sector.}, 
approaches based on flavor symmetry have been proved to be very successful.
One such symmetry named the \mt, originally proposed by Harrison and Scott~\cite{Harrison:2002et}, predicts $ |U_{\mu i }| = |U_{\tau i }|$ (for $i = 1, 2, 3$), where $ U $ represents the Pontecorvo-Maki-Nakagawa-Sakata (PMNS) flavor mixing matrix (see Refs.~\cite{Xing:2015fdg,Xing:2019vks} and the references therein for details).
This symmetry immediately leads to the maximal atmospheric mixing angle $\theta_{23}^{} = \pi/4$ and the Dirac CP-violating phase $\delta = \pm \pi/2$, along with a non-zero $ \theta_{13} $ \footnote{It is worthwhile to note that  the $\mu$-$\tau$ permutation symmetry which was proposed in Ref.~\cite{Fukuyama:1997ky} predicts the maximal atmospheric mixing but with a vanishing $ \theta_{13} $, thus ruled out.}.
It also predicts the trivial Majorana CP-violating phases $\rho, \sigma = 0, \pi/2$, which have totally four different combinations. 
Because of the profound predictability of such symmetry, there exist numerous studies to interpret the leptonic mixing patterns based on the  \mt~\cite{Nishi:2016wki,Zhao:2017yvw,Rodejohann:2017lre,Liu:2017frs,Xing:2017mkx,Xing:2017cwb,Nath:2018fvw,King:2018kka,Nishi:2018vlz,
Nath:2018hjx,Nath:2018zoi,Nath:2018xkz,Huan:2018lzd,Chakraborty:2018dew,Zhu:2018dvj,Zhao:2018vxy,Liao:2019qbb}.
In general, the flavor symmetry is usually imposed at a superhigh energy scale, e.g. $\Lambda_{\mu \tau} \sim 10^{14}~{\rm GeV}$, and needs to be evolved to low energy scales, e.g. $\Lambda_{\rm EW} \sim 10^{2}~{\rm GeV}$, in order to compare with the experimental data.
In this context, the radiative corrections can naturally lead to possible corrections at low energies to the pattern of $(\theta^{}_{23},\delta) = (\pi/4,3\pi/2)$  predicted by the \mt.
Indeed, there have been many theoretical and  phenomenological studies on the breaking of the \mt~induced by the running of renormalization group equations (RGEs)~\cite{Zhao:2017yvw,Rodejohann:2017lre,Liu:2017frs,Xing:2017mkx,Xing:2017cwb,Nath:2018hjx,Huan:2018lzd,Zhu:2018dvj}.
A very interesting possibility has recently been noticed in Ref.~\cite{Huan:2018lzd} that the normal ordering, upper octant of $\theta^{}_{23}$ (i.e. $\theta^{}_{23} > 45^{\circ}$) and the third quadrant of $\delta$ (i.e. $ 180^{\circ} < \delta < 270^{\circ}$) are correlated via RGEs in the framework of minimal supersymmetric standard model (MSSM). 
Motivated by the inspiring T2K measurement, we would like to emphasize this connection and make a more timely and consistent analysis, with all information at hand including oscillation, beta decay, neutrinoless double-beta decay and cosmological observations.

\begin{table}[t]
	\centering  
	\begin{tabular}{c|c|c|c|c|c|c} \hline
		& $ \sin^2 \theta_{23}$ & $\delta/\pi $ &  $\Delta\chi^2_{\rm min}$ \scriptsize(no RGEs) \normalsize&   p-value \scriptsize(no RGEs) \normalsize& $\Delta\chi^2_{\rm min}$ \scriptsize(with RGEs) \normalsize &  p-value \scriptsize(with RGEs) \normalsize\\ \hline 
		T2K ~\cite{Abe:2019vii}   & $0.53^{+0.03}_{-0.04}$    & $1.40^{+0.22}_{-0.18}$  & 1.2& $55\%$ & 0.2 & $90\%$ \\ \hline
		Capozzi \textit{et. al.}~\cite{Capozzi:2020qhw}  & $ 0.545^{+0.01}_{-0.05}$  & $1.28^{+0.38}_{-0.18}$  & 1.1 & $58\%$ & 0.4 & $82\%$\\ \hline
		deSalas \textit{et. al.}~\cite{deSalas:2017kay}  & $ 0.547^{+0.02}_{-0.03}$ & $1.32^{+0.21}_{-0.15}$  & 3.2 & $20\%$ &1.0 & $61\%$ \\  \hline
		Esteban \textit{et. al.}~\cite{Esteban:2018azc}   & $ 0.558^{+0.02}_{-0.033}$    & $1.23^{+0.21}_{-0.16}$  & 4.7 & $10\%$ & 2.3 & $32\%$ \\ \hline
	\end{tabular}
	\caption{ The reported best-fit values along with $ 1\sigma $ errors of $ \sin^2 \theta_{23}$ and $\delta $ from T2K and other global analysis groups. They are to be compared to the $\mu$-$\tau$ symmetry predictions, $ \sin^2 \theta_{23} = 0.5$ and $\delta/\pi = 1.5$. The $\Delta \chi^2_{}$ is defined in Eq.~(\ref{eq:dchis}), which measures the level of deviations from the $\mu$-$\tau$ symmetry predictions to all the relevant experimental data. The p-value is obtained for two degrees of freedom from the corresponding $\Delta \chi^2_{\rm min}$, showing the confidence level to accept the $\mu$-$\tau$ reflection symmetry.}
	\label{tab:Global-fit}
\end{table}

Our procedure to implement the RGE running and the main results are given as follows.
First, the \mt~is assumed at $\Lambda^{}_{\mu\tau}$, which we will fix as the seesaw scale $10^{14}~{\rm GeV}$ \footnote{Varying the seesaw scale is basically equivalent to changing the ratio of Higgs vacuum expectation value $\tan\beta$ in MSSM. Since we let the latter one vary in a very wide range, it is harmless to fix the seesaw scale.}. 
Thus at $\Lambda^{}_{\mu\tau}$, the initial value of $\theta^{}_{23}$ is taken to be $45^{\circ}$ and $\delta = 270^{\circ}$. It is to be noted that the latest T2K measurement and global analyses turn out to disfavor  $\delta = 90^{\circ}$ at $3\sigma$ level~\cite{deSalas:2017kay,Capozzi:2020qhw,Esteban:2018azc}. 
Moreover, the parameter range of $ 0^{\circ} \lesssim \delta \lesssim 180^{\circ}$ ($\sin\delta \gtrsim 0$) is strongly disfavored by the T2K result~\cite{Abe:2019vii}.
Hence, we adopt $\delta = 270^{\circ}$ throughout this work.
Furthermore, there are four different cases for Majorana phases: Case {\bf A}, $\rho = \sigma = 0^{\circ}$; Case {\bf B}, $\rho = \sigma = 90^{\circ}$; Case {\bf C}, $\rho = 0^{\circ}$ and $\sigma = 90^{\circ}$; Case {\bf D}, $\rho = 90^{\circ}$ and $\sigma = 0^{\circ}$. At present, these cases are indistinguishable due to the limited sensitivity of current neutrinoless double-beta decay experiments \cite{Dolinski:2019nrj}. 
After the initial conditions are fixed, for each random choice of parameter set at $\Lambda^{}_{\mu\tau}$ we evolve the system to $\Lambda^{}_{\rm EW}$, and then compare them to the low energy experimental data.
Our primary results are summarized in Table~\ref{tab:Global-fit}, where the p-value signifies the probability for the hypothesis of \mt~to be correct (or alternatively, $1-{\rm p}$ is the probability to reject the \mt). Without radiative corrections, the p-value of \mt~at low energies is only $55\%$ for T2K, while with RGEs in MSSM, this level of acceptance can be improved to as large as $90\%$. For comparison, the results for three global fits are also indicated.
Notably, the global-fit results in Ref.~\cite{Esteban:2018azc} demonstrate that the \mt~is rejected with a $90\%$ C.L., and the inclusion of radiative corrections can reduce it to a more acceptable level of $68\%$.
Further details can be found in Sec.~(\ref{sec:results}).

We outline our paper as follows. In Sec.~(\ref{sec:TheFrame}), we give a brief theoretical description of this work.
Sec.~(\ref{sec:results}) is devoted to our results, where we show that the radiative corrections are carrying $\theta^{}_{23}$ and $\delta$ to their experimentally preferred region by T2K as well as other global fit results. Furthermore, the importance of such breaking on \znbb~ decay has also been discussed. Finally, we summarize in Sec.~(\ref{sec:conclusion}).

\section{Theoretical set-up}\label{sec:TheFrame}
The neutrino mass matrix that obeys the \mt~can be achieved under the following transformations of neutrino fields:
\begin{equation}
\nu_{e,{\rm L}}^{} \leftrightarrow \nu^{\rm c}_{ e, {\rm R}}, ~~~ \nu_{ \mu,{\rm L}}^{} \leftrightarrow \nu^{\rm c}_{\tau, {\rm R}}, ~~~\nu_{ \tau, {\rm L}}^{} \leftrightarrow \nu^{\rm c}_{\mu, {\rm R}} \;,
\end{equation}
where $\nu_{\alpha,{\rm L}}^{}$'s ($\alpha = e,\mu,\tau$) are the left-handed neutrino fields in the flavor basis, and $\nu_{\alpha,{\rm R}}^{\rm c}$'s are the right-handed neutrino charge-conjugated fields. Assuming the neutrino mass term is invariant under such a transformation, the effective Majorana neutrino mass matrix can be read as
\begin{eqnarray}\label{eq:EffMajMass}
M_{\nu} =   \left( \begin{matrix}\langle m\rangle_{ee} & \langle m\rangle_{e\mu} & \langle m\rangle^{\ast}_{e \mu} \cr
\ast & \langle m\rangle_{\mu\mu} & \langle m\rangle^{\ast}_{\mu \tau} \cr
\ast & \ast & \langle m\rangle^{\ast}_{\mu\mu} \cr
\end{matrix} \right) \;.
\end{eqnarray}
It can be noticed from Eq.~(\ref{eq:EffMajMass}) that the different entries of the most general Majorana neutrino mass matrix satisfy the following equalities:
\begin{eqnarray} \label{eq:Mnu_pred}
\langle m\rangle_{ee} = \langle m\rangle_{ee}^* \; , \quad \langle m\rangle_{e\mu}  = \langle m\rangle_{e\tau}^* \; ,\quad \langle m\rangle_{\mu\tau}  = \langle m\rangle_{\mu \tau}^* \; , \quad \langle m\rangle_{\mu\mu}  = \langle m\rangle_{\tau\tau}^* \;.
\end{eqnarray}
The Majorana neutrino mass matrix $M_\nu^{}$ can be diagonalized by the unitary mixing matrix $U$ as $U^{\dagger} M_{\nu}  U^* = M_\nu^{\rm d} = \mathrm{diag}\{m_1^{}, m_2^{}, m_3^{} \}$. Following the standard parameterization advocated by Particle Data Group \cite{Tanabashi:2018oca}, the mixing matrix can be written as 
\begin{align}\label{eq:pmns}
U =  P_l \left(
\begin{matrix}
c^{}_{12} c^{}_{13} & s^{}_{12} c^{}_{13} & s^{}_{13} e^{-{ i} \delta} \cr 
 -s^{}_{12} c^{}_{23} - c^{}_{12} s^{}_{13} s^{}_{23} e^{{ i} \delta} & c^{}_{12} c^{}_{23} -
s^{}_{12} s^{}_{13} s^{}_{23} e^{{ i} \delta} & c^{}_{13}
s^{}_{23} \cr 
 s^{}_{12} s^{}_{23} - c^{}_{12} s^{}_{13} c^{}_{23}
e^{{ i} \delta} & - c^{}_{12} s^{}_{23} - s^{}_{12} s^{}_{13}
c^{}_{23} e^{{ i} \delta} &   c^{}_{13} c^{}_{23} \cr
\end{matrix} \right) P_{\nu}, \;
\end{align}
where $c^{}_{ij} = \cos\theta^{}_{ij} $ and $ s^{}_{ij} = \sin\theta^{}_{ij}$ for $i < j=1, 2, 3$, $ P_l^{} = \mathrm{diag}\{e^{i \phi_{e}},e^{i \phi_{\mu}},e^{i \phi_{\tau}} \}$ contains three unphysical phases which can be absorbed by the rephasing of charged lepton fields, and $ P_{\nu}^{} = \mathrm{diag}\{e^{i \rho},e^{i \sigma},1\}$ is the diagonal Majorana phase matrix. 
Given the \mt, one ends up with following predictions: 
\begin{equation}\label{eq:prediction}
\theta_{23} = 45^\circ,~~~ \delta=\pm 90^\circ,~~~ \rho,~\sigma = 0^{\circ} ~{\rm or}~ 90^\circ,~~~\phi_{e} = 90^\circ,~~~ \phi_{\mu} = - \phi_{\tau}.
\end{equation}
A more detailed discussion for the predictions of \mt~can be found in Ref.~\cite{Nath:2018hjx}.

Having introduced the framework of \mt, in what follows we  describe the 
breaking of such symmetry due to the RGE running effect in the framework of MSSM.
The evolution of neutrino mass matrix $M^{}_\nu$ from $\Lambda^{}_{\rm \mu\tau}$
down to $\Lambda^{}_{\rm EW}$ through the one-loop RGE in  MSSM can be expressed as
\cite{Ellis:1999my,Chankowski:1999xc,Fritzsch:1999ee}
\begin{eqnarray}\label{eq:RelMuTuEW}
M^{}_\nu(\Lambda^{}_{\rm EW})= I^{}_{\nu} I^{\dagger}_{l}
M^{}_{\nu}(\Lambda^{}_{\rm \mu \tau}) I^{*}_{l} \;.
\end{eqnarray}
Here $I^{}_l$ can be approximated as $I^{}_l \simeq \mathrm{diag}\{1, 1, 1-\Delta^{}_{\tau}\}$ together with 
\begin{eqnarray}
I^{}_{\nu} = {\rm exp}\left( \frac{1}{16\pi^2} \int^{\ln (\Lambda^{}_{\rm
EW}/\Lambda)}_{\ln (\Lambda^{}_{\mu \tau}/\Lambda)} \alpha^{}_{\nu} ~ {\rm d}t \right) , \hspace{1cm}
\Delta^{}_{\tau} = \frac{1}{16\pi^2} \int^{\ln( \Lambda^{}_{\mu \tau}/\Lambda)}_{\ln (\Lambda^{}_{\rm
EW}/\Lambda)}y^2_{\tau}~{\rm d}t \;,
\end{eqnarray}
where $t \equiv \ln(\mu/\Lambda)$ with $\mu$ being the running energy scale and $\Lambda$ being an arbitrary cutoff, and $\alpha^{}_{\nu} \simeq -6/5 g^2_{1} - 6 g^2_{2} + 6 y^{2}_{t} $ with $g^{}_{1}$ and $g^{}_{2}$ being the gauge couplings and $y^{}_{t}$ being the Yukawa coupling of the top quark.
Note that the Yukawa coupling of $\tau$ lepton is boosted by the ratio of Higgs vacuum expectation value (vev) $\tan\beta$ in MSSM through $y^{2}_\tau = (1+ \tan^2{\beta}) m^2_\tau/v^2$ with the Higgs vev being $ v \simeq 174 $ GeV.
The strength of the RGE corrections can be greatly enhanced if a large value of $\tan\beta$ has been taken. Note that to avoid the quark Yukawa couplings being enhanced to the non-perturbative region, the ratio of Higgs vev should be bounded with $\tan\beta \lesssim 50$. 
Since the experimental knowledge of supersymmetry and the values of  $\tan\beta$ are currently lacking, in this note we shall vary  $\tan\beta$ freely from $10$ to $50$.

The leptonic parameters at a given energy scale can be obtained by diagonalizing the mass matrix $M^{}_\nu$.
We define $\Delta \theta^{}_{ij} \equiv \theta^{}_{ij}
\left(\Lambda_{\rm EW}^{}\right) - \theta^{}_{ij}\left(\Lambda_{\mu\tau}^{}\right)$
(for $i < j = 1, 2, 3$), $\Delta \delta \equiv \delta\left(\Lambda_{\rm EW}^{}\right) - \delta\left(\Lambda_{\mu\tau}^{}\right)$,
$\Delta \rho \equiv \rho \left(\Lambda_{\rm EW}^{}\right)
 -\rho\left(\Lambda_{\mu\tau}^{}\right)$ and
$\Delta \sigma \equiv \sigma\left(\Lambda_{\rm EW}^{}\right) -\sigma\left(\Lambda_{\mu\tau}^{}\right)$ to measure the strengths of RGE-induced
corrections to the leptonic mixing parameters.
To the leading order approximation, the three light neutrino
masses at $\Lambda^{}_{\rm EW}$  are expressed as~\cite{Huan:2018lzd}
\begin{align}\label{eq:RGEMass}
m_1(\Lambda_{\rm EW}) & \simeq m_1(\Lambda_{\mu \tau})[1 - \Delta_{\tau}(1-c^{2}_{12}c^{2}_{13})]I^{2}_\nu \;, \nonumber \\ 
m_2(\Lambda_{\rm EW}) & \simeq m_2(\Lambda_{\mu \tau})[1 - \Delta_{\tau}(1-s^{2}_{12}c^{2}_{13})]I^{2}_\nu \;, \nonumber \\
m_3(\Lambda_{\rm EW}) & \simeq m_3(\Lambda_{\mu \tau})[1 - \Delta_{\tau}c^{2}_{13}]I^{2}_\nu \;.
\end{align}
Here $\theta^{}_{12}$ and $\theta^{}_{13}$ take their values at $\Lambda^{}_{\rm EW}$.
%
The leptonic flavor mixing angles at low energies are given by
\begin{eqnarray} \label{eq:RGEAngles}
\Delta \theta_{12}^{}  & \simeq & \frac{\Delta_{\tau}^{}}{2} c_{12}^{}
s_{12}^{} \left[s_{13}^2\left(\zeta_{31}^{\eta_{\rho}^{}}
- \zeta_{32}^{\eta_{\sigma}^{}} \right)
+ c_{13}^2 \zeta_{21}^{-\eta_{\rho}^{} \eta_{\sigma}^{}} \right] \;,
\nonumber\\
\Delta \theta_{13}^{}  & \simeq & \frac{\Delta_{\tau}^{}}{2} c_{13}^{}
s_{13}^{} \left(c_{12}^2 \zeta_{31}^{\eta_{\rho}^{}}
+ s_{12}^2 \zeta_{32}^{\eta_{\sigma}^{}}\right) \;,
\nonumber\\
\Delta \theta_{23}^{}  & \simeq & \frac{\Delta_{\tau}^{}}{2} \left(s_{12}^2
\zeta_{31}^{-\eta_{\rho}^{}} + c_{12}^2
\zeta_{32}^{-\eta_{\sigma}^{}}\right) \;.
\end{eqnarray}
Finally,  one can calculate three CP-violating phases at low energies as
\begin{eqnarray}\label{eq:RGEPhases}
\Delta \delta  & \simeq &  \frac{\Delta_{\tau}^{}}{2}
\left[\frac{c_{12}^{} s_{12}^{}}{s_{13}^{}}
\left(\zeta_{32}^{-\eta_{\sigma}^{}} -
\zeta_{31}^{-\eta_{\rho}^{}}\right) -
\frac{s_{13}^{}}{c_{12}^{} s_{12}^{}}
\left(c_{12}^4 \zeta_{32}^{-\eta_{\sigma}} -
s_{12}^4 \zeta_{31}^{-\eta_{\rho}^{}} +
\zeta_{21}^{\eta_{\rho}^{} \eta_{\sigma}}\right)\right] \;,
\nonumber\\
\Delta \rho & \simeq & \Delta_{\tau}^{}
\frac{c_{12}^{} s_{13}^{}}{s_{12}^{}} \left[ s_{12}^{2}
\left(\zeta_{31}^{-\eta_{\rho}} -
\zeta_{32}^{-\eta_{\sigma}^{}}\right) +\frac{1}{2}
\left(\zeta_{32}^{-\eta_{\sigma}^{}} +
\zeta_{21}^{\eta_{\rho}^{} \eta_{\sigma}}\right)\right] \;,
\nonumber \\
\Delta \sigma & \simeq & \Delta_{\tau}^{}
\frac{s_{12}^{} s_{13}^{}}{2 c_{12}^{}}
\left[ s_{12}^2 \left(\zeta_{21}^{\eta_{\rho}^{}\eta^{}_\sigma}
- \zeta_{31}^{-\eta_{\rho}^{}}\right) -
c_{12}^{2} \left(2 \zeta_{32}^{-\eta_{\sigma}^{}} -
\zeta_{31}^{-\eta_{\rho}^{}} - \zeta_{21}^{
	\eta_{\rho}^{} \eta_{\sigma}^{}}\right)\right] \;,
\end{eqnarray}
where $\eta_{\rho}^{} \equiv \cos 2\rho = \pm 1$ and $\eta_{\sigma}^{} \equiv
\cos 2\sigma = \pm 1$ represent different choices of $\rho$ and $\sigma$ at $\Lambda^{}_{\mu\tau}$,
and the ratios $\zeta^{}_{ij} \equiv (m^{}_i - m^{}_j)/
(m^{}_i + m^{}_j)$ are defined with $m^{}_i$ and $m^{}_j$ at $\Lambda^{}_{\rm EW}$
(for $i, j = 1, 2, 3$).
In obtaining the above equations, the \mt, i.e., $\theta^{}_{23}(\Lambda_{\mu\tau}^{}) = 45^\circ$ and $\delta(\Lambda_{\mu\tau}^{}) = -90^\circ$ have also been applied. 
We now proceed to present a detailed description of our numerical results in  subsequent sections. Note that instead of the approximated formalism in this section we will solve the exact one-loop RGEs for the numerical calculations.

\section{Results}\label{sec:results}
In this section we present a close-up of the low energy predictions of \mt~in the MSSM framework and examine their impacts on \znbb~decays. To cover up the observables at $\Lambda^{}_{\rm EW}$ within their experimentally allowed ranges, we need to scan each free inputs at $\Lambda^{}_{\mu\tau}$ with wide enough ranges, and run their values to $\Lambda^{}_{\rm EW}$  via the one-loop RGEs.
Then we compare the output parameters at $\Lambda^{}_{\rm EW}$ to the low energy experimental knowledge.
For this purpose, a chi-square function has been defined as
\begin{eqnarray} \label{eq:dchis}
\Delta\chi^2 = \Delta\chi^2_{\rm osc} + \Delta\chi^2_{\rm \beta} + \Delta\chi^2_{\rm \beta\beta} + \Delta\chi^2_{\rm cosmo} + \chi^2_{\mu\tau}\;,
\end{eqnarray}
where $\chi^2_{\rm osc}$ stands for the oscillation data, $\chi^2_{\rm \beta}$ for the beta decays, $\chi^2_{\rm \beta\beta}$ for the $0\nu\beta\beta$ decays, $\chi^2_{\rm cosmo}$ for the cosmological observations, and $\chi^2_{\mu\tau}$ defined in Eq.~(\ref{eq:mutau-chisq}) measures the distance of the observed $\theta^{}_{23}$ and $\delta$ to their $\mu$-$\tau$ symmetry predictions.
Therefore the expectation of $\Delta\chi^2$ should be close to 0 if the hypothesis of $\mu$-$\tau$ reflection symmetry is correct.
The deviation of $\Delta\chi^2$ from 0 demonstrates the significance to reject our model.
The details are summarized as follows.
For  $\chi^2_{\rm osc}$, the global-fit results of neutrino oscillation parameters including two mass squared differences and two mixing angles $\theta^{}_{12}$ and $\theta^{}_{13}$ from Ref.~\cite{Capozzi:2020qhw} will be used. The oscillation chi-square is constructed with the standard procedure using the provided central values and symmetrized one standard deviations. 
The data of non-oscillation experiments mainly constrain the absolute scale of neutrino masses at $\Lambda^{}_{\rm EW}$.
For $\chi^2_{\rm \beta}$, the beta-decay limits of $m^{}_{\beta}$ from Mainz~\cite{Kraus:2004zw}, Troitsk~\cite{Aseev:2011dq} as well as the latest release of KATRIN~\cite{Aker:2019uuj, Aker:2019qfn} are adopted. The results of Troitsk and Mainz are given as $m^{2}_{\beta} = -0.67 \pm 2.53~{\rm eV}^2$ and $m^{2}_{\beta} = -0.6 \pm 3.0~{\rm eV}^2$, respectively. As for KATRIN, the approximated likelihood $\mathcal{L}^{}_{\rm KATRIN}$ with a skewed normal distribution can be found in Ref.~\cite{Huang:2019tdh}, and we transform this likelihood to the chi-square function according to the relation $\chi^2 = -2 \ln \mathcal{L}$.
Following Ref.~\cite{Capozzi:2020qhw}, we construct our $\chi^2_{\rm \beta\beta}$ by transforming the $90\%$ C.L. combined upper limit on the effective neutrino mass, $|m^{}_{ee}| < 110~{\rm meV}$~\cite{Agostini:2019hzm}, into $|m^{}_{ee}| = 0 \pm 0.07~{\rm eV}$.
The cosmological likelihood for the sum of three neutrino masses is obtained by analyzing the Markov chain files provided by the Planck Legacy Archive~\cite{Aghanim:2018eyx}.
To be specific, we choose the the dataset of \textit{Planck} TT, TE, EE + lowE + lensing + BAO, which leads to $\Sigma = m^{}_{1} + m^{}_{2} +m^{}_{2} < 0.12~{\rm eV} $ at $95\%$ C.L. Again, this likelihood is converted to $\chi^2_{\rm cosmo}$ with $\chi^2 = -2 \ln \mathcal{L}$. 
To find how far the oscillation data are deviating from the $\mu$-$\tau$ symmetry scenario, it is convenient to define the following chi-square function
\begin{eqnarray}\label{eq:mutau-chisq}
\chi^2_{\mu\tau} = \frac{\left[\sin^2\widetilde{\theta}^{}_{23} - \sin^2\theta^{}_{23}(\Lambda^{}_{\rm EW})\right]^2}{(\sigma^{-}_{\sin^2\theta_{23}})^2} + \frac{\left[\widetilde{\delta}^{} - \delta^{}(\Lambda^{}_{\rm EW})\right]^2}{(\sigma^{+}_{\delta})^2} \;,
\end{eqnarray}
where $\widetilde{\theta}^{}_{23}$ and $\widetilde{\delta}$ take their experimental best-fit values, $\theta^{}_{23}(\Lambda^{}_{\rm EW})$ and $\delta(\Lambda^{}_{\rm EW})$ are the low energy predictions of the $\mu$-$\tau$ reflection symmetry. 
Since the best fit indicates $\theta^{}_{23} > 45^{\circ}$ and $\delta < 270^{\circ}$, we adopt one side of the unsymmetric reported errors, e.g. $\sigma^{-}_{\sin^2\theta_{23}} = 0.04$ and $\sigma^{+}_{\delta} = 0.22 \pi$ for T2K.

\begin{figure*}[t!]
	\begin{center}
		\hspace{-0.1cm}
		\subfigure{\includegraphics[width=0.48\columnwidth]{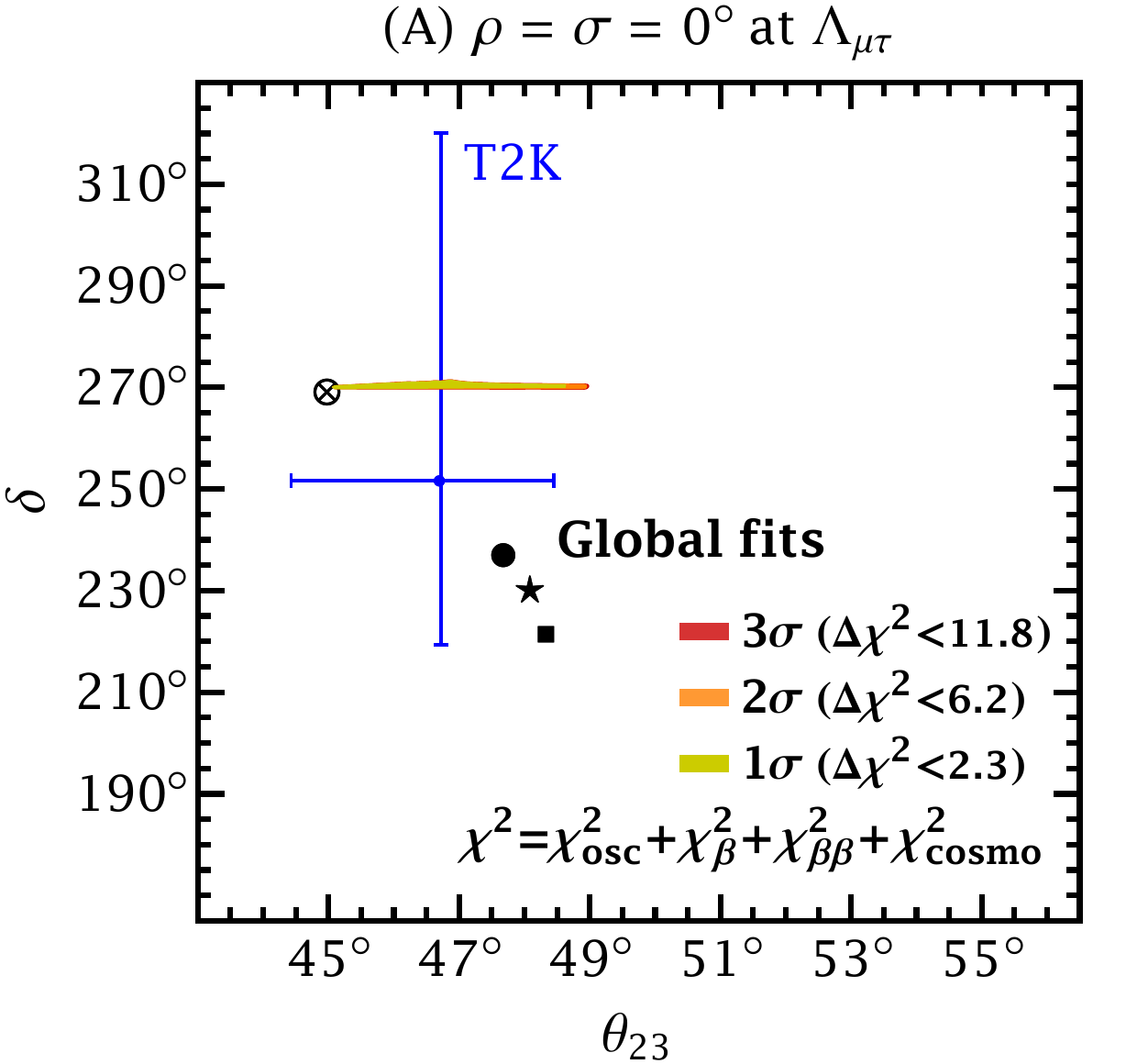}} 	
		\subfigure{\includegraphics[width=0.48\columnwidth]{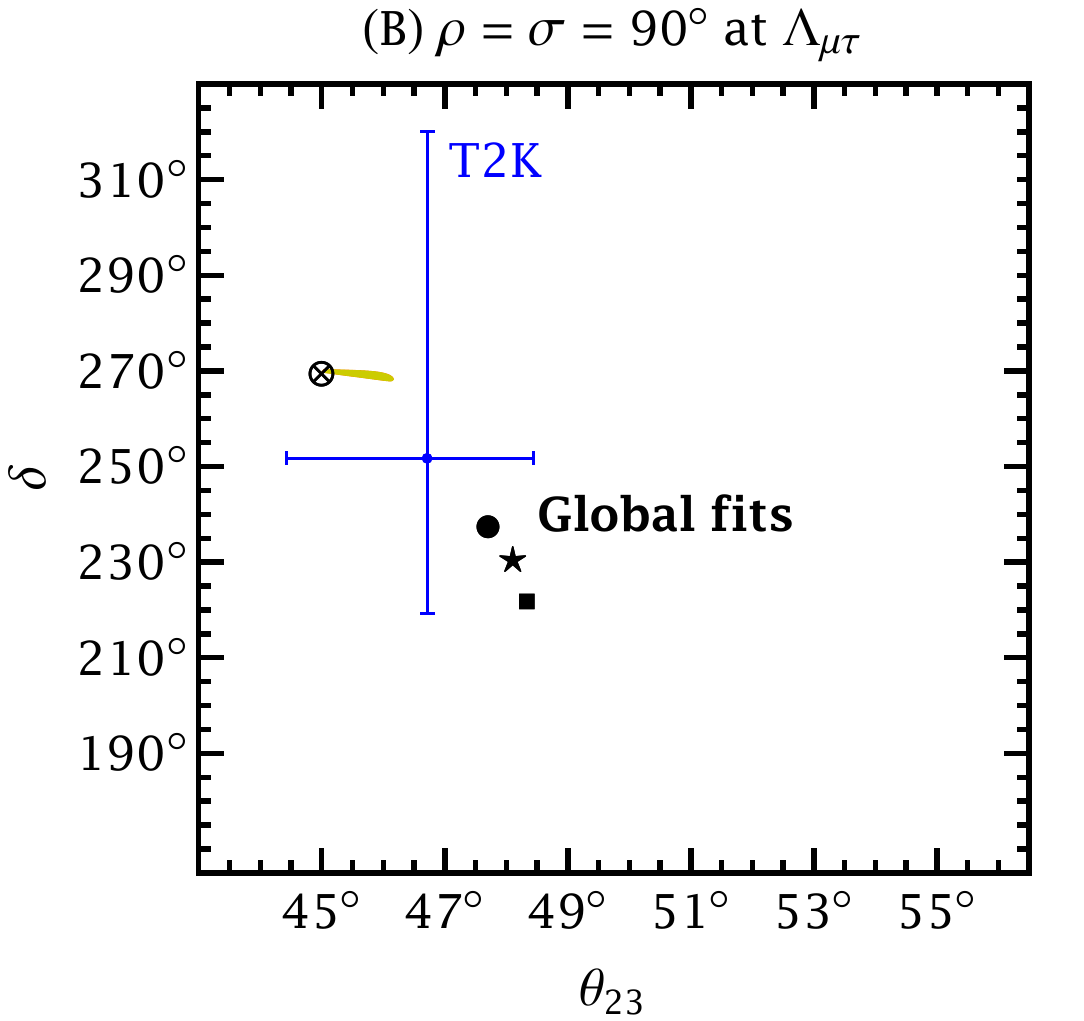}}
		\subfigure{\includegraphics[width=0.48\columnwidth]{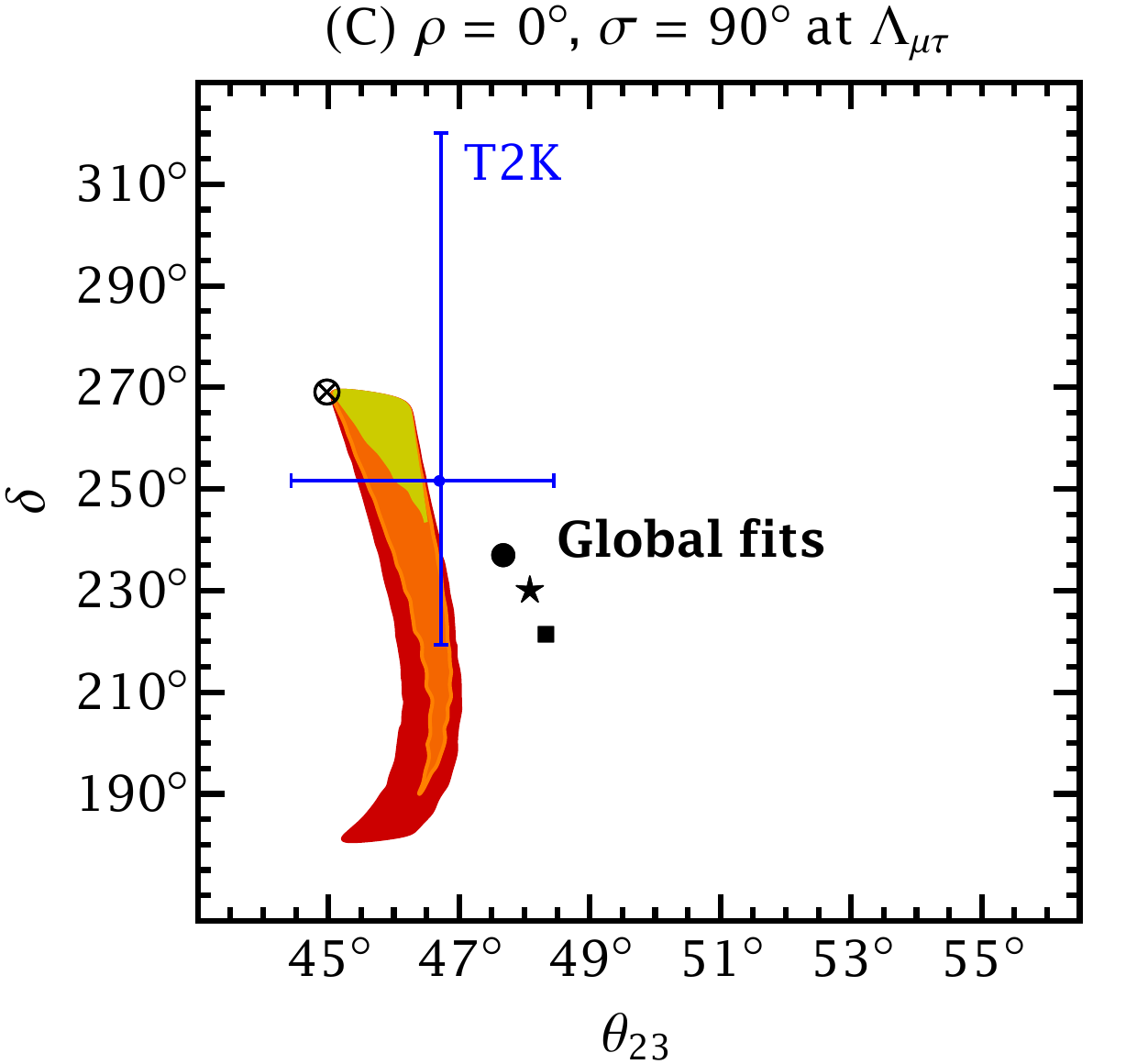}}
		\subfigure{\includegraphics[width=0.48\columnwidth]{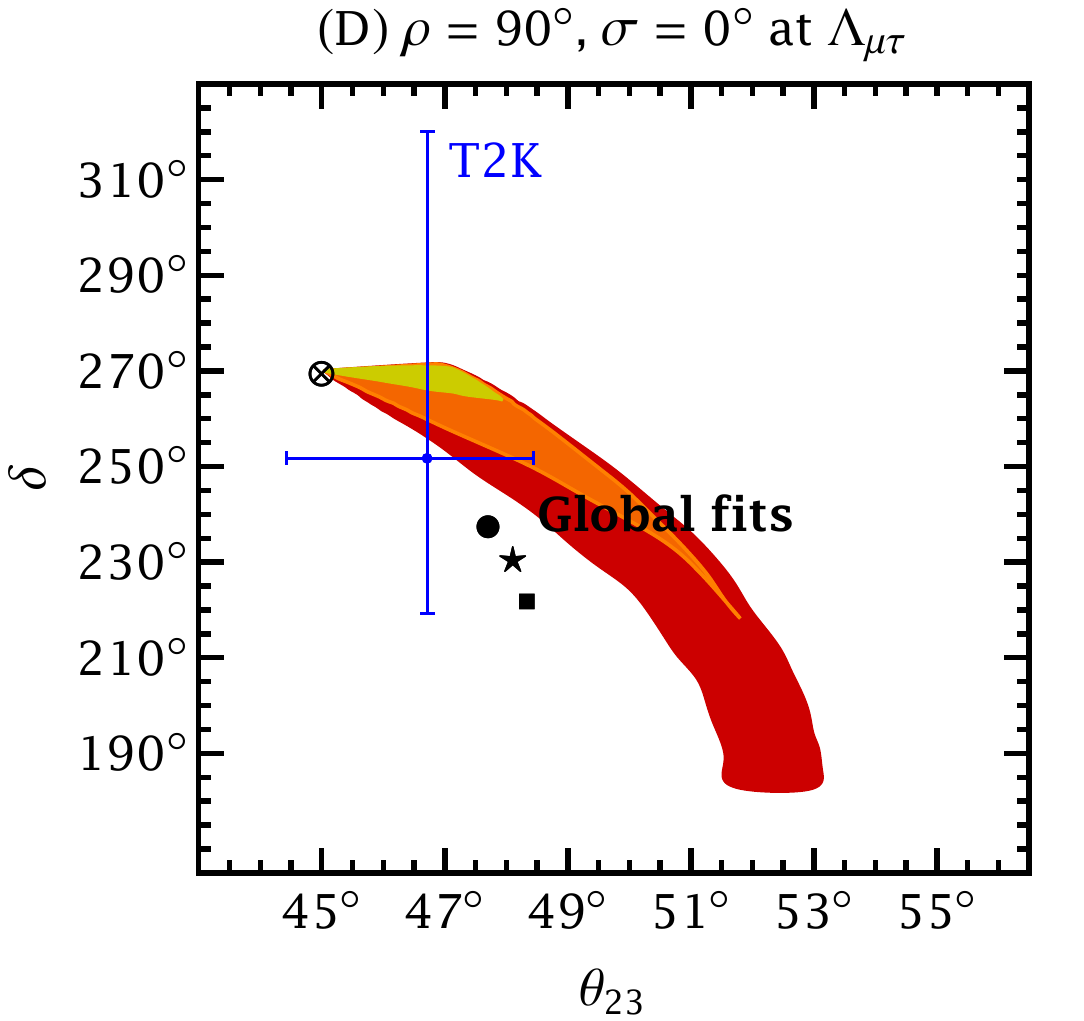}}
	\end{center}
	\vspace{-0.8cm}
	\caption{ The predictions of $\theta^{}_{23}$ and $\delta$ at $\Lambda_{\rm EW}^{}$  compared to the global-fit results~\cite{Capozzi:2020qhw}. 
		Four different cases for Majorana phases are considered: Case {\bf A}, $\rho = \sigma = 0^{\circ}$ (\emph{Top-left Panel}); Case {\bf B}, $\rho = \sigma = 90^{\circ}$ (\emph{Top-right Panel}); Case {\bf C}, $\rho = 0^{\circ}$ and $\sigma = 90^{\circ}$ (\emph{Bottom-left Panel}); Case {\bf D}, $\rho = 90^{\circ}$ and $\sigma = 0^{\circ}$ (\emph{Bottom-right Panel}).	
		The ratio of Higgs vevs in MSSM $\tan{\beta}$ has been marginalized over $[10\cdots 50]$. The black cross stands for the $\mu$-$\tau$ symmetry point $(\theta^{}_{23},\delta) = (45^{\circ},270^{\circ})$. The yellow, orange and red regions are predicted by the RGE-induced $\mu$-$\tau$ reflection symmetry breaking with $1\sigma$, $2\sigma$ and $3\sigma$ levels, respectively. 
		The vertical and horizontal error bars show the $1\sigma$ uncertainty of the T2K measurement.	
		The best-fit values of three global analyses are marked as the filled circle (Capozzi \textit{et. al.}~\cite{Capozzi:2020qhw}), star (deSalas \textit{et. al.}~\cite{deSalas:2017kay}) and square (Esteban \textit{et. al.}~\cite{Esteban:2018azc}). Have in mind that the uncertainties of the current global fits are comparable to T2K.}
	\label{fig:t23delta}
\end{figure*}

{Our target is to compare the model with the \mt~(alternative hypothesis, reduced model) to the free model without such symmetry (null hypothesis, full model). 
The reduced model can be simply obtained by fixing some of the free parameters in the full model.
A $\chi^2$ difference test needs to be performed to decide between these two models.
For this purpose, we need to compare the minima of $\chi^{2}_{}$ obtained with and without the \mt .
The degrees of freedom $df$ for this test are the difference of respective degrees of freedom  for these two models, i.e. $df=df^{}_{\rm reduced}-df^{}_{\rm full}$.
For a statistical test of a single model, the degrees of freedom are calculated by subtracting number of the random variables (e.g. $\Sigma$, $m^{}_{\beta}$, $|m^{}_{ee}|$, $\Delta m^2_{21}$, $\Delta m^2_{31}$, $\theta^{}_{12}$, $\theta^{}_{13}$, $\theta^{}_{23}$ and $\delta$ at $\Lambda^{}_{\rm EW}$) by that of the estimated model parameters (e.g. $m^{}_{1}$, $\Delta m^2_{21}$, $\Delta m^2_{31}$, $\theta^{}_{12}$, $\theta^{}_{13}$ and $\tan\beta$ at $\Lambda^{}_{\mu\tau}$). 
The difference of degrees of freedom in our test is $df=2$, contributed by the fixed values of $\theta^{}_{23}$ and $\delta$ in the reduced model with the \mt . Note that with the current experimental precision, two Majorana phases $\rho$ and $\sigma$ are far from being measured (minimizing them does not reduce the minimum value of $\chi^2$ at all); therefore they do not contribute to the additional reduction of degrees of freedom for the full model. The p-value thus should be derived from $\chi^2$ with two degrees of freedom in our test.
}

Now we are ready to perform the numerical calculation with the one-loop RGEs. To efficiently scan the parameter space and sample the minimum of $\chi^2$, the \texttt{MultiNest}~\cite{Feroz:2007kg,Feroz:2008xx,Feroz:2013hea} routine will be adopted. In Table~\ref{tab:Global-fit}, we have shown the minimum of $\Delta\chi^2$ with(out) the radiative corrections for different experimental inputs of $\theta^{}_{23}$ and $\delta$. The corresponding p-value has been indicated, which gives a direct numerical measure of how likely the $\mu$-$\tau$ symmetry can be accepted, while its complementary ($1-{\rm p}$) measures the significance that the $\mu$-$\tau$ reflection symmetry is rejected. We notice that T2K is yielding data favoring the $\mu$-$\tau$ reflection symmetry in comparison to earlier global analysis~\cite{deSalas:2017kay,Esteban:2018azc}, which is also reflected in the latest global analysis~\cite{Capozzi:2020qhw} with the T2K results included. In all cases, taking into account the radiative corrections in MSSM can well improve the fit.
Next, we will explicitly demonstrate how the RGEs can lead $\theta^{}_{23} = 45^{\circ}$ and $\delta = 270^{\circ}$ at $\Lambda^{}_{\mu\tau}$ to their experimental favored values at $\Lambda^{}_{\rm EW}$.
%

\begin{figure*}[t!]
	\begin{center}
		\hspace{-0.1cm}
		\subfigure{\includegraphics[width=0.48\columnwidth]{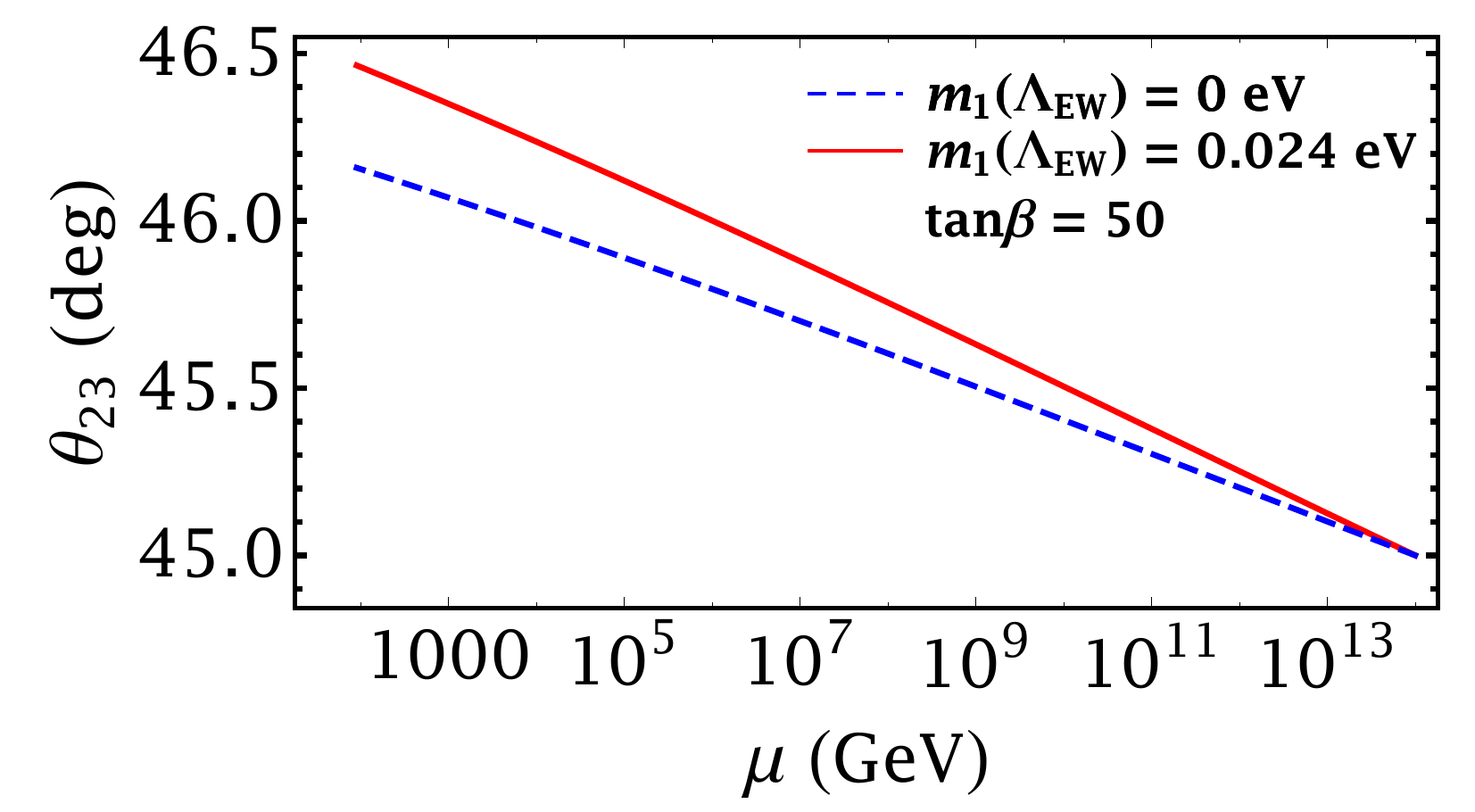}}
		\subfigure{\includegraphics[width=0.48\columnwidth]{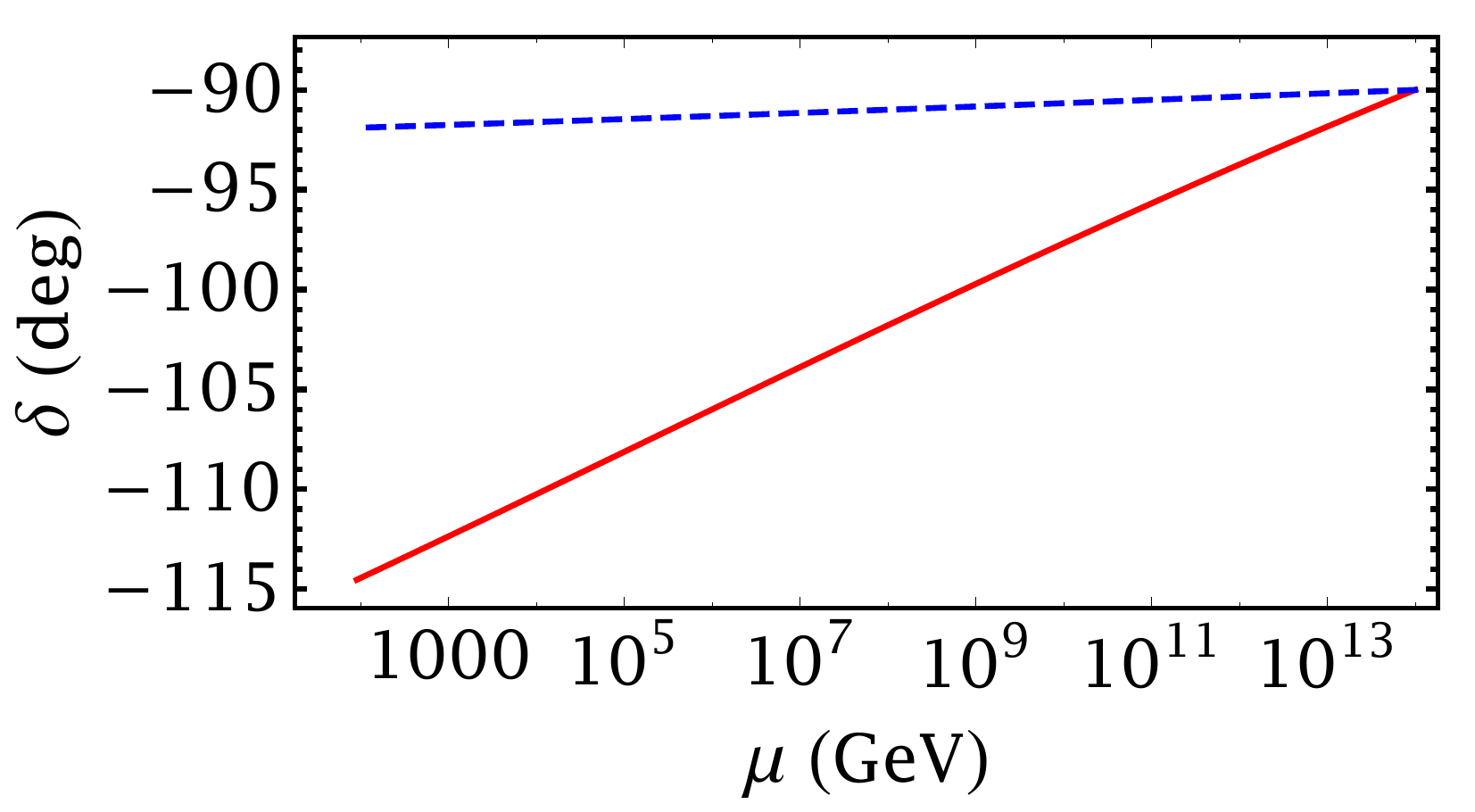}}
		\subfigure{\includegraphics[width=0.48\columnwidth]{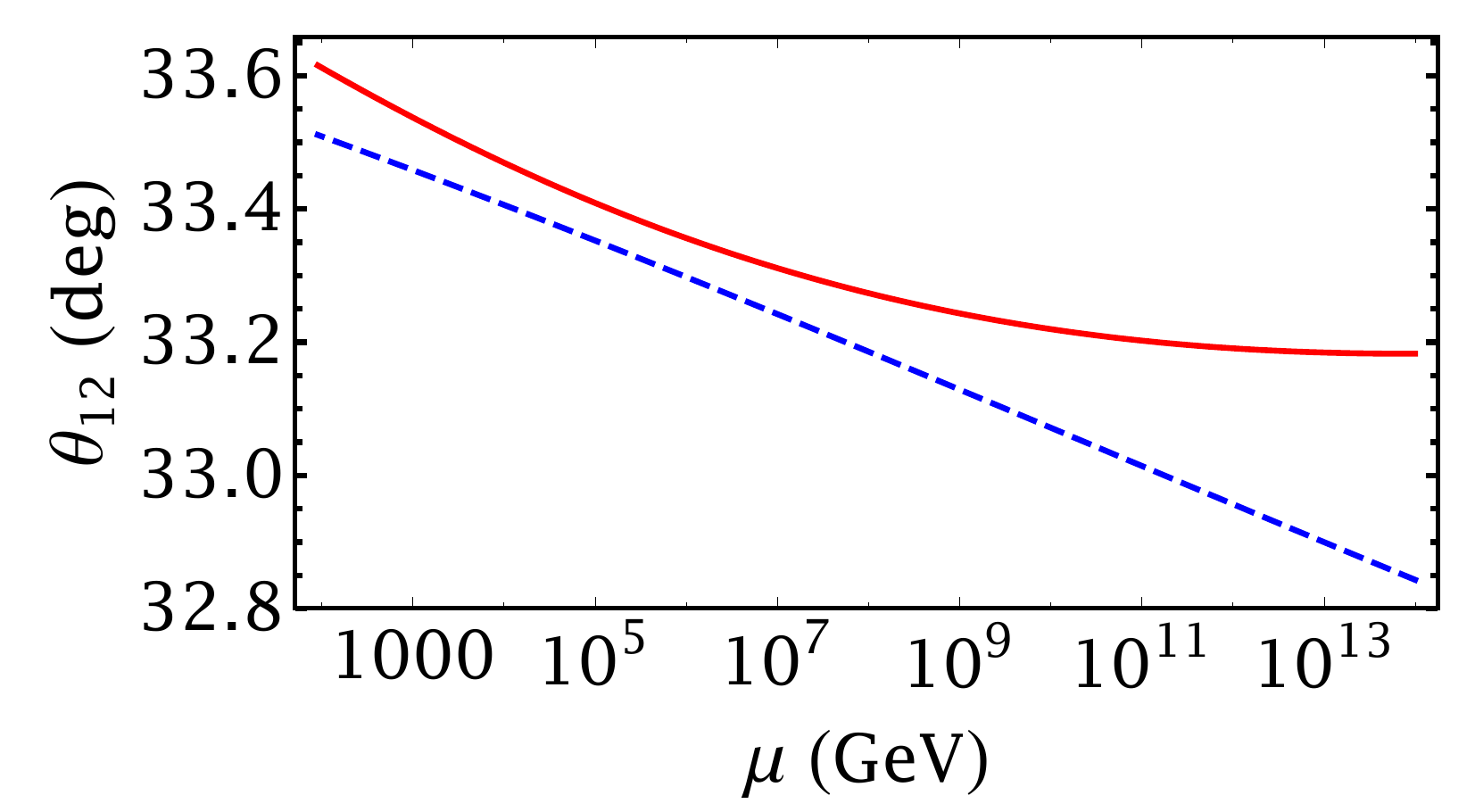}} 	
		\subfigure{\includegraphics[width=0.48\columnwidth]{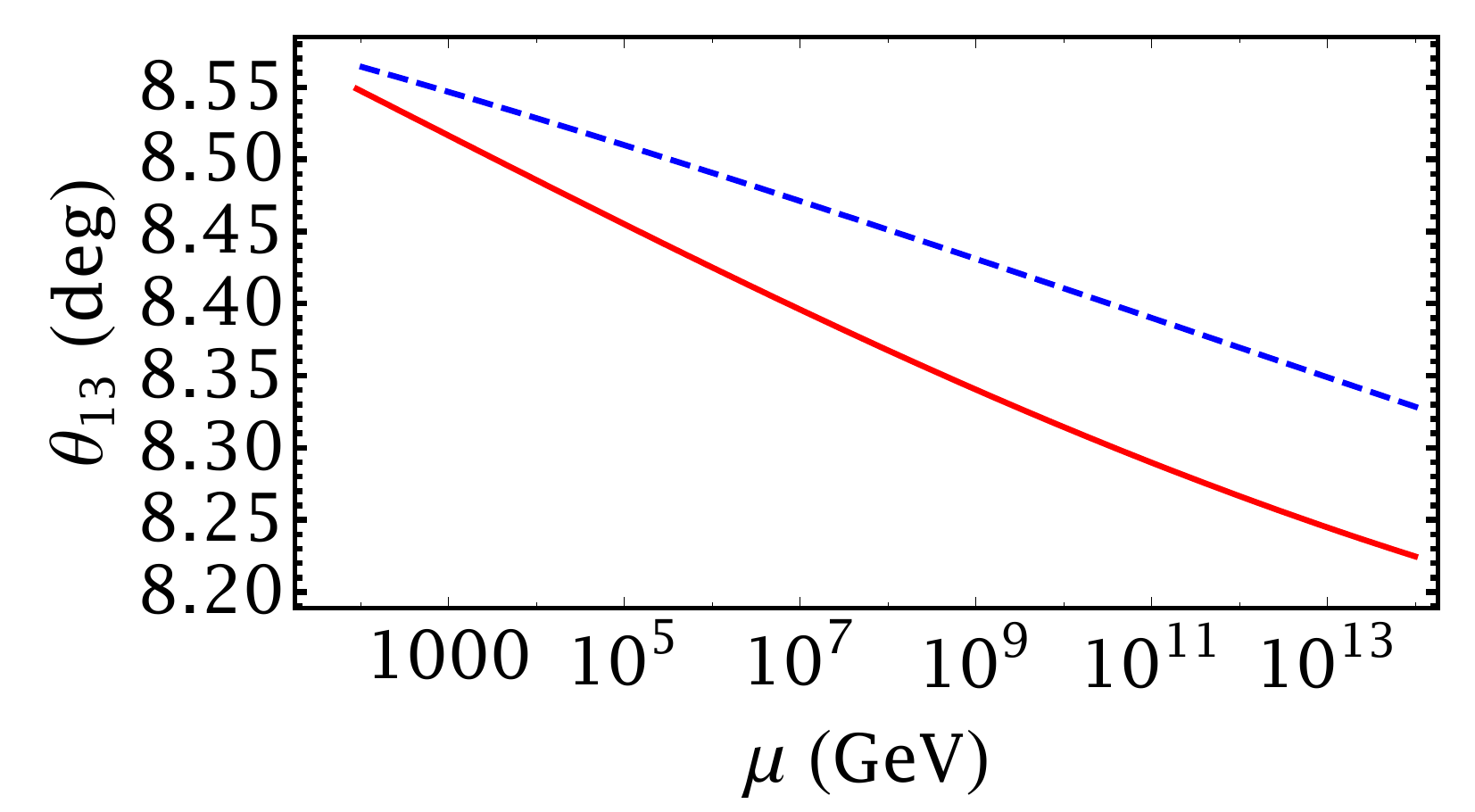}}
		\subfigure{\includegraphics[width=0.48\columnwidth]{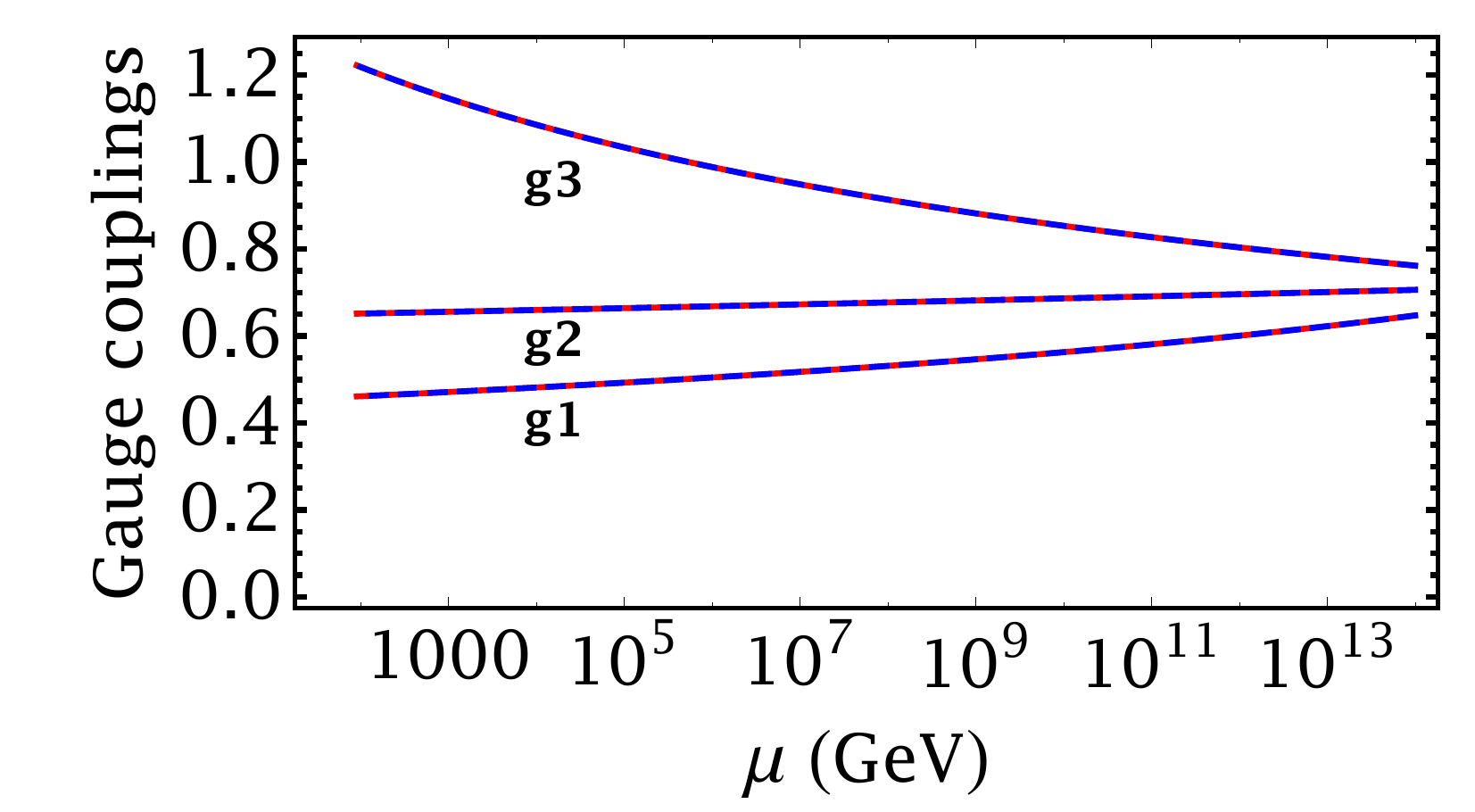}}
		\subfigure{\includegraphics[width=0.48\columnwidth]{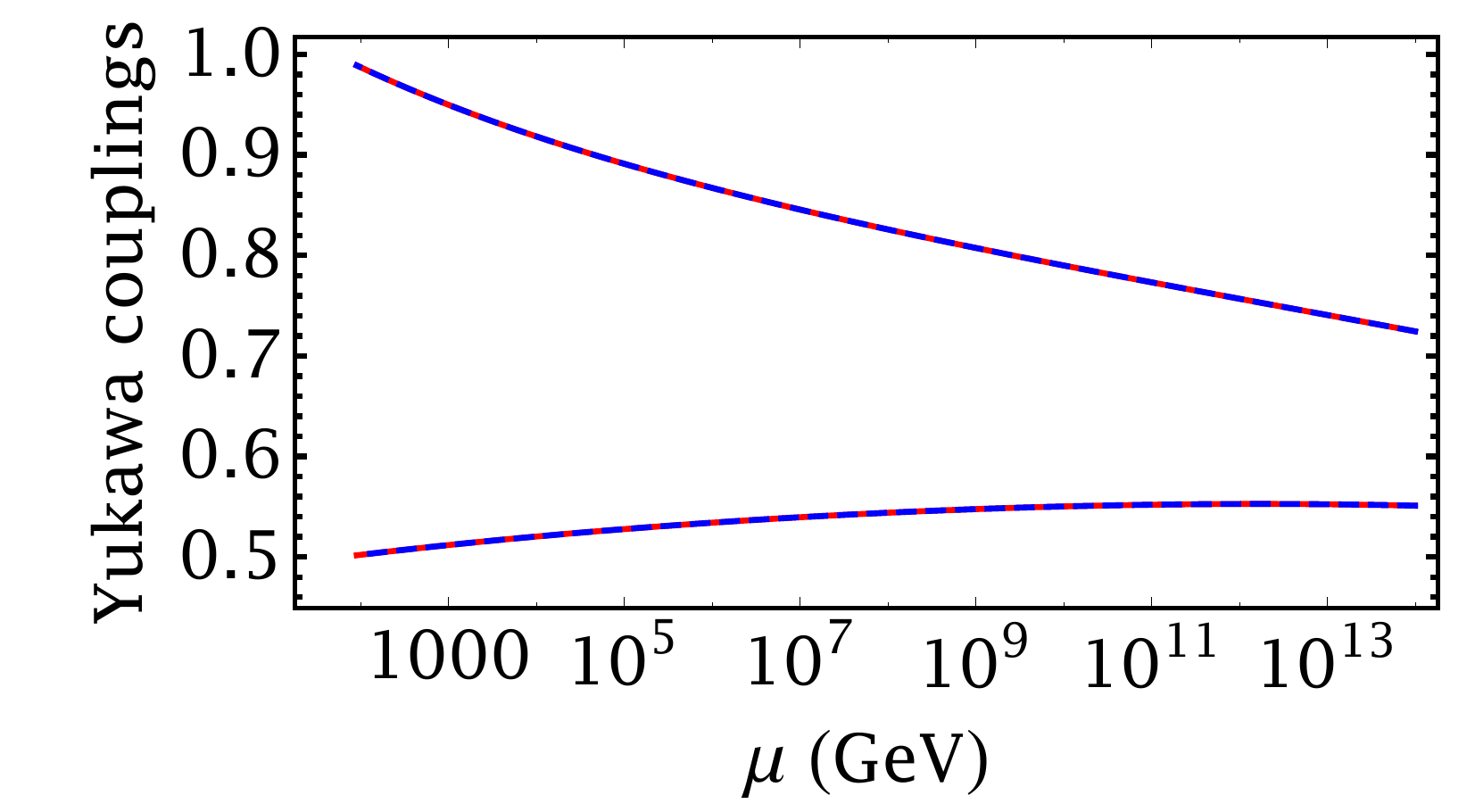}}
	\end{center}
	\vspace{-0.8cm}
	\caption{The RGE running of leptonic mixing parameters $\theta^{}_{12}$, $\theta^{}_{13}$, $\theta^{}_{23}$ and $\delta$, the gauge couplings $g^{}_{1}$, $g^{}_{2}$ and $g^{}_{3}$, and the Yukawa couplings $y^{}_{t}$ and $y^{}_{\tau}$ from the energy scale $\mu = 10^{14}~{\rm GeV}$ to $\mu = M^{}_{Z}$. The $\mu$-$\tau$ symmetry has been assumed at $\Lambda^{}_{\mu\tau} = 10^{14}~{\rm GeV}$. The initial values of Majorana phases have been taken as $\rho = 0^{\circ}$ and $\sigma = 90^{\circ}$. The dashed blue curves and the red ones stand for different cases with $\chi^2 = 0.2$ and $\chi^2 = 2.2$, respectively.}
	\label{fig:running}
\end{figure*}
Fig.~\ref{fig:t23delta} shows the low energy correlation between $\theta^{}_{23}$ and $\delta$ due to the RGE-triggered \mt~breaking.
Four different panels represent four possible combinations of  Majorana phases as predicted by the $\mu$-$\tau$ reflection symmetry.
The shaded regions for $1\sigma$ (yellow), $2\sigma$ (orange), and $3\sigma$ (red) C.L. are obtained by requiring $\Delta \chi^2$ to be smaller than certain values as indicated in the figure. Note that we have removed temporarily $\chi^2_{\mu\tau}$ from the total $\chi^2$ in obtaining the shaded region, in order to have a clearer parameter comparison.
The horizontal and vertical  blue error bars stand for the T2K $1\sigma$ results for $\theta^{}_{23}$ and $\delta$, and their cross represents the best-fit point. Three global-fit results are shown in the figure as the filled circle~\cite{Capozzi:2020qhw}, star~\cite{deSalas:2017kay} and square~\cite{Esteban:2018azc}, respectively.
From all the panels, we notice that due to the RGE-induced breaking, mixing angle $ \theta_{23} $ tends to favor higher octant at low energies. This is in excellent agreement with the latest T2K results as well as with the latest global-fit data.  However, the amount of deviations are very different for different initial values of $ \rho, \sigma $ at $ \Lambda_{\mu \tau} $. 
It is worth mentioning that the numerical results obtained here are in good agreement with the analytical results.
To understand this breaking pattern, we notice  from the third line of Eq.~(\ref{eq:RGEAngles}) that  the $ \mathcal{O}(\Delta_{\tau}) $ term depends on different CP-violating factors (i.e., $ \zeta^{-\eta_{\rho}}_{31}$ and $ \zeta^{-\eta_{\sigma}}_{32} $), which adds a very distinct contribution to $  \theta_{23}  $ depending on the initial choices of $ \rho$ and $\sigma $ at $ \Lambda_{\mu \tau} $. 
On the other hand, the RGE-running behavior of $ \delta $ also shows very different patterns depending on $ \rho$ and $\sigma $. 
One can observe deviations of $ \delta $ from the maximal CP violation less than $ \mathcal{O}(5^\circ) $  in the top two panels, whereas deviations as large as $ \mathcal{O}(90^\circ) $ can be noticed from the bottom two panels.
As we require $\Delta \chi^2$ to be smaller (e.g., from $3\sigma$ to $1\sigma$), the allowed regions shrink significantly.
This is due to that the general size of the radiative correction is depending on the value of $m^{}_{1}$, which is severely constrained by the Planck result. 
%
%

{Furthermore, as an explicit example, in Fig.~\ref{fig:running} we show the RGE-running behavior of different mixing parameters $\theta^{}_{12}$, $\theta^{}_{13}$, $\theta^{}_{23}$ and $\delta$,  the gauge couplings $g^{}_{1}$, $g^{}_{2}$ and $g^{}_{3}$, and the Yukawa couplings $y^{}_{t}$ and $y^{}_{\tau}$. We choose two scenarios for the initial conditions, corresponding to the dashed blue curves and red ones in Fig.~\ref{fig:running} respectively. Some comments are given as follows:
\begin{itemize}
		\item For the first scenario, shown in dashed blue curves, 
		the initial values of parameters at $\Lambda^{}_{\mu\tau} = 10^{14}~{\rm GeV}$ have been taken as $\tan\beta = 50$, $y^{}_{t} = 0.724$, $y^{}_{\tau}=0.551$, $m^{}_{1} = 0~{\rm eV}$, $\Delta m^{2}_{\rm sol} = 1.168 \times 10^{-4}~{\rm eV^2}$, $\Delta m^{2}_{\rm atm} = 3.994 \times 10^{-3}~{\rm eV^2}$, $\theta^{}_{12} = 32.84^{\circ}$ and $\theta^{}_{13}= 8.33^{\circ}$, which leads to the best-fit case $\chi^2_{\rm min} = 0.2$. 
		The initial value of the lightest neutrino mass $m^{}_{1}$ for the best fit is found to be zero. This preference is mainly driven by the experimental results of beta decay, neutrinoless double-beta decay and cosmological observations. 
		For this best-fit case, we have $ \theta^{}_{23} = 46.2^\circ $ and $\delta = -92^\circ$ at $\Lambda_{\rm EW}$.
		The RGE correction to $\delta$ is tiny with a vanishing $m^{}_{1}$ for $\rho = 0^{\circ}$ and $\sigma = 90^{\circ}$, such that $\delta$ at low energy does not develop a significant deviation with its initial value $-90^{\circ}$ at $\Lambda^{}_{\mu\tau}$. Even though the T2K experiment prefers $\delta = -108^{\circ}$, the uncertainty of the measurement is still relatively large, and $\delta$ with small RGE corrections can still have an acceptable fit with the data.
		\item For the second scenario in red curves, the parameters with different initial values at $\Lambda^{}_{\mu\tau} = 10^{14}~{\rm GeV}$ are $m^{}_{1} = 0.029~{\rm eV}$, $\Delta m^{2}_{\rm sol} = 1.47 \times 10^{-4}~{\rm eV^2}$, $\Delta m^{2}_{\rm atm} = 4.002 \times 10^{-3}~{\rm eV^2}$, $\theta^{}_{12} = 33.18^{\circ}$ and $\theta^{}_{13}= 8.22^{\circ}$, which gives a larger value of $\chi^2 = 2.2$. Even though the oscillation parameters in this case agree better with the experimental data in comparison to the best-fit case. The choice of a bigger $m^{}_{1}$ to generate large correction to $\delta$ increases the global $\chi^2$ defined in Eq.~(\ref{eq:dchis}).
\end{itemize}
It can be seen from the second row that deviations less than $\mathcal{O}(1^\circ)$ have been identified for both $\theta^{}_{12}$ and $ \theta^{}_{13}$ at $\Lambda_{\rm EW}$, which fall within the $1\sigma$ confidence level of the latest global analysis of neutrino oscillation data as given by Ref.~\cite{Capozzi:2020qhw}. 
We show the RGE-running behavior for the gauge and Yukawa couplings at the third row. Note that the Yukawa coupling of $\tau$-lepton is greatly enhanced by $\tan\beta = 50$.
}

\begin{figure}[t!]
	\centering
	\hspace{-1.cm}
	\subfigure{\includegraphics[width=0.45\columnwidth]{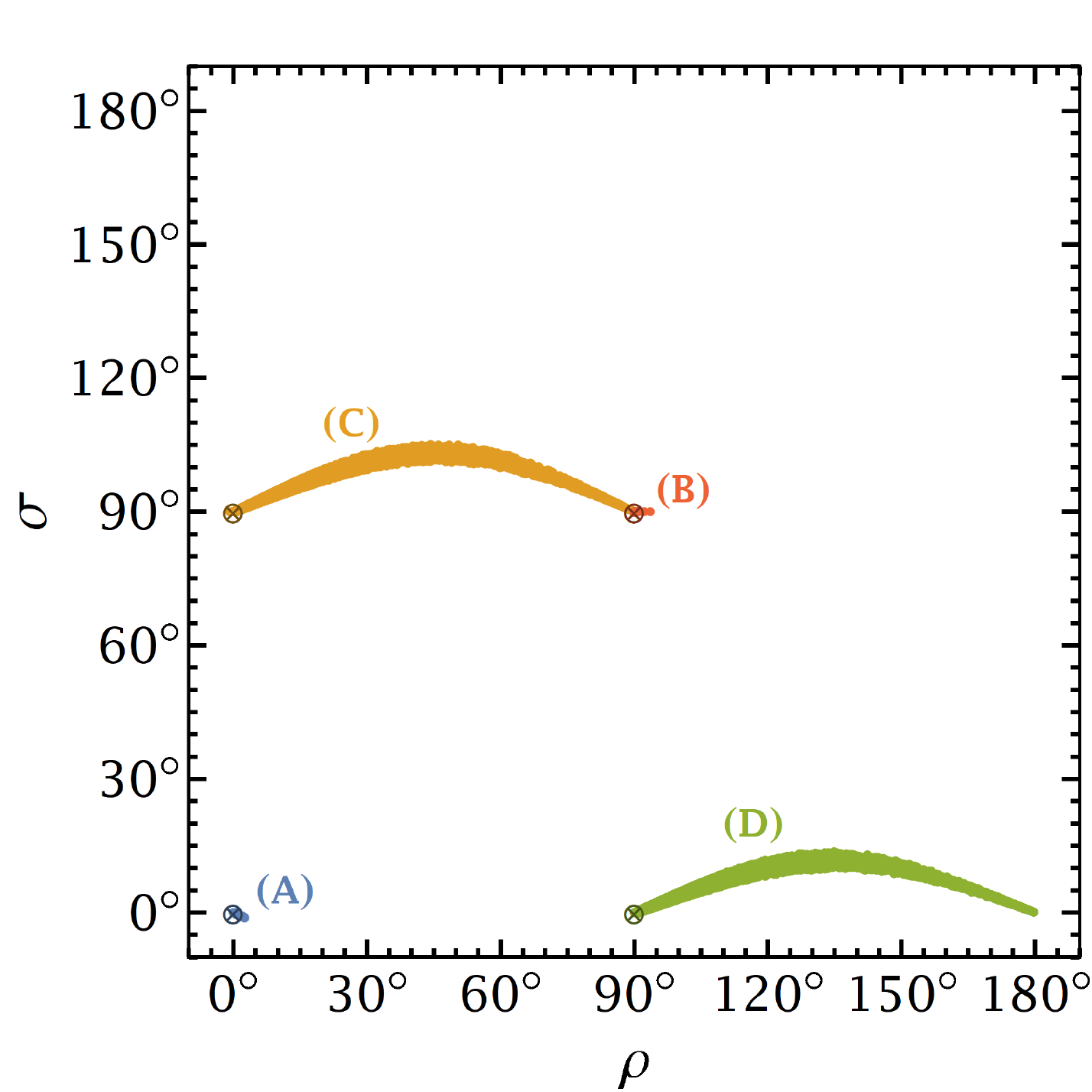}}
	\hspace{0.5cm} 	
	\subfigure{\includegraphics[width=0.45\columnwidth]{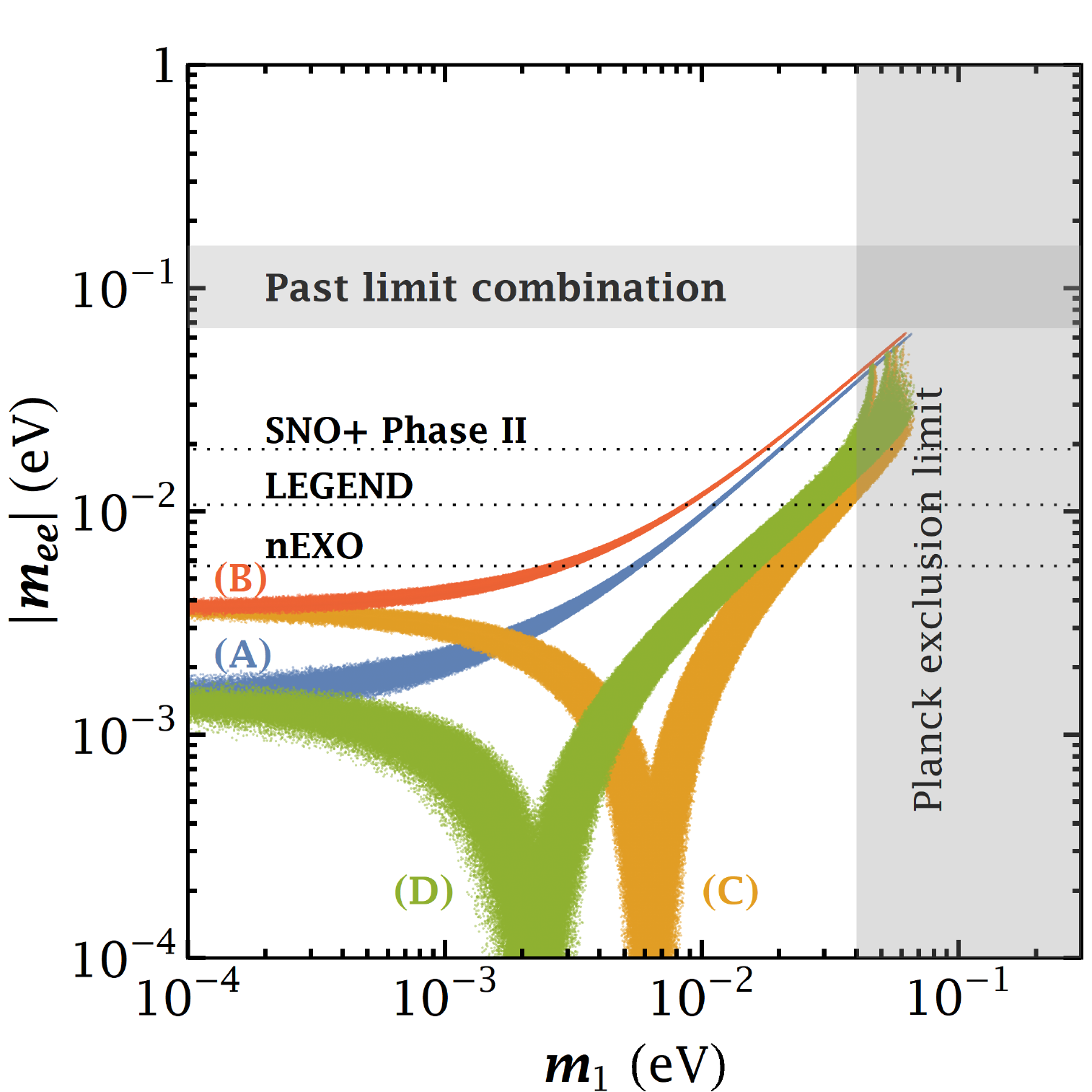}} 	
	\caption{\emph{Left Panel}: the Majorana phase $\sigma$ versus $\rho$. \emph{Right Panel}: the effective Majorana neutrino mass $| m^{}_{ee}|$ as a function of the lightest neutrino mass $m^{}_{1}$ in normal ordering. The shaded colorful regions are predicted from a RGE-induced broken $\mu$-$\tau$ reflection symmetry at the $3\sigma$ level.
		Four different cases for Majorana phases are considered: Case {\bf A}, $\rho = \sigma = 0^{\circ}$; Case {\bf B}, $\rho = \sigma = 90^{\circ}$; Case {\bf C}, $\rho = 0^{\circ}$ and $\sigma = 90^{\circ}$; Case {\bf D}, $\rho = 90^{\circ}$ and $\sigma = 0^{\circ}$.
		In the right panel, the current experimental limit on $| m^{}_{ee}|$ is shown by the horizontal gray band, whereas the projected sensitivities of future experiments are indicated by the dotted horizontal lines (see text for more details).
		The latest Planck bound on $m^{}_{1}$ is shown by the vertical gray band  corresponding to $ \Sigma < 0.12~{\rm eV}$ at the 95\% C.L.} 
	\label{fig:mee}
\end{figure}


In the following, we examine the impact of the RGE-induced symmetry breaking on \znbb~decays and confront them with the future experimental sensitivities~\cite{KamLAND-Zen:2016pfg,Alduino:2017ehq,Albert:2017owj,Agostini:2018tnm,Andringa:2015tza,
Abgrall:2017syy,Albert:2017hjq,Azzolini:2020skx}. 
Our results can be summarized in Fig.~\ref{fig:mee} for four different cases of Majorana phases. 
The shaded regions are generated by allowing $\Delta \chi^2 < 11.83$, which corresponds to the $3\sigma$ level of confidence.
We notice from the left panel that RGEs tend to take $\rho$ and $\sigma$ in Cases {\bf C} and {\bf D} to their high energy values in Cases {\bf B} and {\bf A}, respectively. 
For the right panel,
one can observe that the cancellation in $|m^{}_{ee}|$ still exists for Cases {\bf A} and {\bf B}. In contrast, there are clear lower bounds for Cases {\bf C} and {\bf D}, i.e., $|m^{}_{ee}| \gtrsim 3~{\rm meV}$ and $|m^{}_{ee}| \gtrsim 1~{\rm meV}$ at $3\sigma$ C.L. respectively. These two cases can be completely excluded if a future ultimate sensitivity of $|m^{}_{ee}| = 1~{\rm meV}$ is achieved \cite{Cao:2019hli}.
For comparison,
the horizontal gray band indicates the combined limits of previous experiments as in Ref.~\cite{Agostini:2019hzm}, and the limits on $|m^{}_{ee}|$ can be read as $(66 - 155\; \mathrm{meV})$.
Moreover, the most optimistic sensitivities projected for SNO+ Phase II $(19 - 46 \; \mathrm{meV})$ \cite{Andringa:2015tza},  LEGEND $(10.7 - 22.8\; \mathrm{meV})$ \cite{Abgrall:2017syy}, and nEXO $(5.7 - 17.7\; \mathrm{meV})$ \cite{Albert:2017hjq}
are shown by the dotted lines.
On the other hand, the vertical gray band represents the current limit of cosmological data on the sum of three neutrino masses $ \Sigma < 0.12~{\rm eV}$ from the Planck collaboration~\cite{Aghanim:2018eyx}.

\section{Conclusion}\label{sec:conclusion}
We have made an attempt to give a theoretical explanation of the latest  T2K results based on a flavor symmetry with rich phenomenology, the \mt. 
An immediate predictions of such symmetry are $ \sin^2\theta_{23} = 0.5$ and $ \delta = 0.5 \pi$ or $1.5 \pi$  along with the trivial values of the Majorana CP phases. 
On the other hand, the best-fit values of  T2K data turn out to be $\sin^2 \theta_{23} = 0.53$ and $\delta = 1.4\pi$ for normal ordering, deviating from the \mt~predictions by around $1\sigma$ level. The global-fit results can yield more severe tensions.
We find that the RGE-induced \mt~within the MSSM framework can successfully explain such slight deviations. 
Our main results are presented in Table~\ref{tab:Global-fit} and Fig.~\ref{fig:t23delta}.
For instance, the level to accept the \mt~can be improved from $55\%$ to $90\%$ for the inputs of T2K.
It can be noticed from Fig.~\ref{fig:t23delta} that for Cases {\bf C} and {\bf D} both $\theta_{23}$ and $\delta$ are in excellent agreement with the T2K measurements.
Likewise, we have discussed the consequences of the RGE-induced \mt~breaking in \znbb~decays.
A future sensitivity of $|m^{}_{ee}| \simeq 3~{\rm meV}$ can completely rule out Case {\bf B} even with RGE corrections in MSSM, while a sensitivity of $|m^{}_{ee}|\simeq 1~{\rm meV}$ is required to exclude Case {\bf A}.

\acknowledgements 
	{Authors would like to thank Prof.~Zhi-zhong Xing and Prof.~Shun Zhou for inspiring discussions and reading the manuscript. Authors also thank Dr.~Jing-yu Zhu for helpful discussions.
	NN is grateful to Dr. Eduardo Peinado for his suggestion to address the latest T2K results.
	 GYH is supported in part by the National Natural Science Foundation of China under grant No.~11775232 and No.~11835013. NN is supported by the postdoctoral fellowship program DGAPA-UNAM, CONACYT CB-2017-2018/A1-S-13051 (M\'exico) and DGAPA-PAPIIT IN107118. }
\noindent

%


\bibliographystyle{utphys}
\bibliography{bibilography}
\end{document}